\documentclass[aps,prl,twocolumn,showpacs,superscriptaddress]{revtex4-1}
\usepackage[dvips]{color}
\usepackage{graphicx}
\usepackage{bm}% bold math

\begin{document}

\title{Shell plus pairing effect arguments for cluster preformation at the 
nuclear surface in cold fission}

\author{D. N. Poenaru$^*$ and R. A. Gherghescu}
\email[]{poenaru@fias.uni-frankfurt.de}
%\homepage[]{http://fias.uni-frankfurt.de/{\verb+~+{poenaru}}
\affiliation{
Horia Hulubei National Institute of Physics and Nuclear
Engineering (IFIN-HH), \\P.O. Box MG-6, RO-077125 Bucharest-Magurele,
Romania and\\ Frankfurt Institute for Advanced Studies, Johann Wolfgang
Goethe University, Ruth-Moufang-Str. 1, D-60438 Frankfurt am Main, Germany}

\date{ }

\begin{abstract}

In 1928 G. Gamow as well as Condon and Gurney gave the first explanation of
alpha decay as a quantum tunnelling of a preformed particle at the nuclear
surface.  Soon after experimental discovery in 1984 by Rose and Jones of
cluster radioactivity, confirming earlier (1980) predictions by Sandulescu,
Poenaru and W.  Greiner, a microscopic theory also explained the phenomenon
in a similar way.  Here we show for the first time that in a spontaneous
cold fission process the shell plus pairing corrections calculated with
Strutinsky's procedure may give a strong argument for preformation of a
light fission fragment near the nuclear surface.  It is obtained when the
radius of the light fragment, $R_2$, is increased linearly with the
separation distance, $R$, of the two fragments, while for $R_2=$~constant
one gets the well known two hump potential barrier.

\end{abstract}

\pacs{25.85.Ca, 24.75.+i, 21.10.Tg, 27.90.+b}
%25.85.Ca Spontaneous fission
%24.75.+i General properties of fission
%21.10.Tg Lifetimes
%27.90.+b properties of nuclei with A (greater-than-or-equal-to) 220

\maketitle

In 1928 G. Gamow \cite{gam28zp} as well as Condon and Gurney \cite{gur28n}
gave the first explanation of alpha decay as a quantum tunnelling of a
preformed particle at the nuclear surface.  Soon after experimental
discovery in 1984 by Rose and Jones \cite{ros84n} of cluster radioactivity,
confirming earlier (1980) predictions by Sandulescu, Poenaru and W.  Greiner
\cite{enc95}, a microscopic theory \cite{lov98pr} also explained the
phenomenon in a similar way.  Here we show for the first time that in a
spontaneous cold fission process \cite{sig81jpl} the shell plus pairing
corrections calculated with Strutinsky's procedure \cite{str67np} may give a
strong argument for preformation of a light fission fragment near the
nuclear surface.  It is obtained when the radius of the light fragment,
$R_2$, is increased linearly with the separation distance, $R$, of the two
fragments, while for $R_2=$~constant one gets the well known two hump
potential barrier.

Among the almost 2450 nuclides (nuclear species) known up to now only 288
are stable to occur primordially: their half-lives are comparable to, or
longer than the Earth's age (4.5 billion years), hence a significant amount
survived since the formation of the Solar System.  The metastable nuclides
are decaying toward the stable ones.

The first informations about nuclei have been obtained in 1896, when Antoine
Henri Becquerel discovered a ``mysterious'' radiation of a uranium salt
(potassium uranyl sulfate) which was bent by a magnetic field.  The term
radioactivity was coined by Marie Curie.  Together with her husband, Pierre
Curie, they discovered radium (symbol Ra) and polonium (Po), which possess a
million times much stronger radioactivity.

Ernest Rutherford (ER) gave the names $\alpha$~($^4$He nuclei),
$\beta$~(electrons) and $\gamma$~(electromagnetic radiation with frequencies
of $3 \times 10^{19}$ Hz or higher and wavelengths of $10^{-11}$~m (10~pm)
or lower).  Three of the four fundamental forces (strong, weak, and
electro-magnetic) are responsible for them.  They are produced by the
decay of excited nuclei of radioactive elements.  Gamma rays can penetrate
through several centimeters of lead and large doses of them are harmful. 
From scattering experiments (1911) ER deduced that atomic particles
consisted primarily of empty space surrounding a central core called
nucleus.  He transmuted one element into another, elucidated the concepts of
the half-life and decay constant.  By bombarding nitrogen with
$\alpha$-particles produced oxygen.  The atomic nucleus was discovered
around 1911.

After 1928 the microscopic theories of $\alpha$~decay have been developed,
see e.g.  \cite{lan60rmp}.  The theory was also extended to explain cluster
decays \cite{lov98pr}.  Simple relationships are also very useful
\cite{p83jpl80,wan15prc} to estimate the half-lives.

The liquid drop model (LDM) was introduced by Sir John William Strutt, Lord
Rayleigh.  His book {\em Theory of Sound}, was published in 1878.  Niels
Bohr applied the LDM to Nuclear Physics \cite{boh36n}.  It was used by Lise
Meitner and her nephew O.R.  Frisch \cite{mei39n} to explain the {\it
induced fission} discovered by Otto Hahn and Fritz Strassmann, who
identified by chemical means among the fission fragments a barium isotope
\cite{hah39n}.  N.  Bohr and J.A.  Wheeler \cite{boh39pr} published a
theoretical paper, based on his LDM; they showed that fission was more
likely to occur with $^{235}$U than $^{238}$U.  An interesting historical
account of the discovery of induced fission was written in 1984 by the
famous Edoardo Amaldi, former member of the Enrico Fermi's (EF) team who did
the first experiment two years before Otto Hahn, but was wrong in
interpreting the data.  Even a genius like EF was a human being, hence could
be sometimes wrong.  The energetic (nuclear power plants) and military
applications of the induced fission changed completely our world.

Spontaneous fission was discovered in 1940 by G.N. Flerov and K.A. Petrzhak
\cite{fle40pr}.  Usually the fission fragments are deformed and excited;
they decay by neutron emission and/or $\gamma$~rays, so that the total
kinetic energy (TKE) of the fragments is smaller by about 25-35 MeV than the
released energy, or Q-value.  The asymmetric mass distributions of the
fission fragments and the spontaneously fissioning shape isomers
\cite{pol62sjetp} could not be explained until 1967, when V.M.  Strutinsky
reported \cite{str67np} his macroscopic-microscopic method.  His
calculations gave for the first time a two hump potential barrier.  Shape
isomers occupied the second minimum.

Besides $\alpha$, $\beta$, $\gamma$ decay and fission there are other types
of nuclear disintegrations sometimes referred to ``exotic decay modes'' as
beta-delayed particle emissions, particle-accompanied fission (or ternary
fission), fissioning shape-isomers, proton radioactivity, heavy particle
radioactivities (HPR) \cite{p140b89,p195b96}, etc.  A brief presentation, at
a level of non-specialist, of the large diversity of nuclear decay modes may
be found in the Ref.~\cite{pg267ecmp}.
\begin{figure}
\centerline{\includegraphics[width=8cm]{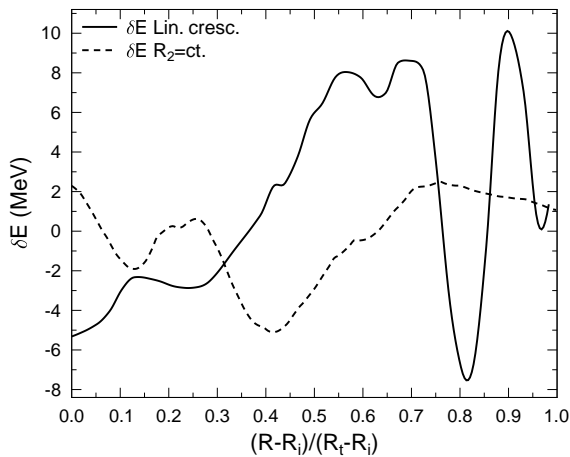}} %str282cn00lcbw}}
\caption{Comparison of absolute values of shell and pairing
correction energies for symmetrical fission of $^{282}$Cn with
$R_2$~constant (dashed line) and linearly increasing $R_2$ (solid line). 
\label{shpe}} 
\end{figure} 
\begin{figure} 
\centerline{\includegraphics[width=8cm]{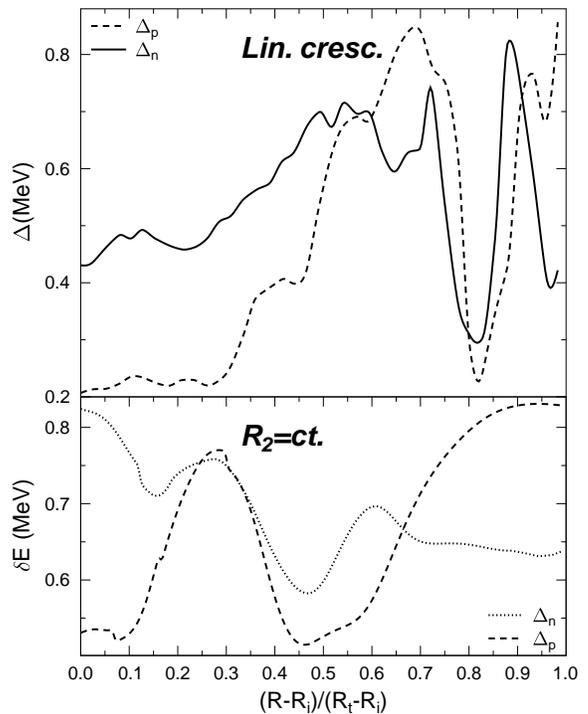}} %lamdel282cn00lc}}
\caption{Solutions of BCS equations for symmetrical fission of $^{282}$Cn
with linearly increasing $R_2$ (top) and constant $R_2$ (bottom). The gap
for protons and neutrons do have a similar behaviour with that of the shell
corrections.
\label{bcs}} 
\end{figure}

Superheavy nuclei with atomic numbers $Z=104-118$ are produced by fusion
reactions \cite{ham13arnps,oga01n}.  The simplest way to identify a new
superheavy element synthesized in such a way is to measure its
$\alpha$~decay chain, down to a known nuclide.  Sometimes this is not
possible since its main decay mode could be spontaneous fission.  For atomic
numbers larger than 121 cluster decay may compete as well \cite{p309prl11}. 
Among the many theoretical papers in this field one should mention
\cite{sob11ra} and \cite{sta13prc,war11prc}.

We reported \cite{p348prc16} results obtained within macroscopic-microscopic
method using cranking inertia \cite{bra72rmp} and the best two-center shell
model \cite{ghe03prc} in the plane of two independent variables $(R,\eta)$,
where $R$ is the separation distance of the fragments and $\eta =(A_1 -
A_2)/A$ is the mass asymmetry with $A, A_1, A_2$ the mass numbers of the
parent and nuclear fragments.  Phenomenological deformation energy,
$E_{Y+E}$, was given by Yukawa-plus-exponential model \cite{kra79pr}, and
the shell plus pairing corrections, $\delta E = \delta U + \delta P$ are
based on the asymmetric two center shell model (ATCSM).  This time we give
more detailed arguments for the neighboring nucleus $^{282}$Cn.  The deep
minimum of total deformation energy near the surface is shown for the first
time as a strong argument for cluster preformation.

An outline of the model was presented previously \cite{p348prc16}. Here we
repeat just few lines.  The parent $^A Z$ is split in two fragments: the
light, $^{A_2}Z_2$, and the heavy one, $^{A_1}Z_1$ with conservation of
hadron numbers $A=A_1+A_2$ and $Z=Z_1+Z_2$.  The corresponding radii are
given by $R_0=r_0 A^{1/3}$, $R_{2f}=r_0 A_2^{1/3}$, and $R_{1f}=r_0
A_1^{1/3}$ with $r_0=1.16$~fm.  The separation distance of the fragments is
initially $R_i = R_0$ and at the touching point it is $R_t = R_{1f} +
R_{2f}$.  The geometry for linearly increasing $R_2$ from 0 to $R_{2f}=R_e$
is defined by: \begin{equation} R_2 = R_{2f}\frac{R-R_i}{R_t-R_i}
\end{equation}

According to the macroscopic-microscopic method the total deformation energy
contains the macroscopic Yukawa-plus-exponential (Y+EM) term and the shell
plus pairing corrections 
\begin{equation} 
E_{def} = E_{Y+E} + \delta E
\end{equation}
In units of $\hbar \omega_0^0 = 41 A^{-1/3}$ the shell corrections are
calculated with the Strutinsky procedure as a sum of protons and neutrons
contributions
\begin{equation}
\delta u = \delta u_p + \delta u_n
\end{equation}
One obtains a minimum when there are important bunchings of levels (high
degeneracy of the quantum state: the same energy corresponds to several
states).

The BCS \cite{bar57pr} theory was first introduced in condensed matter in
order to explain the superconductivity at a very low temperature.  It was
extended to nuclei for explanation of the pairing interaction, see e.g. 
\cite{bra72rmp}.  By solving the BCS system of two equations, with two
unknowns, we find the Fermi energy, $\lambda$, and the pairing gap $\Delta$,
separately for protons and neutrons.  The total pairing corrections are
given by 
\begin{equation} 
\delta p = \delta p_p + \delta p_n 
\end{equation}
and finally the total shell plus pairing corrections in MeV 
\begin{equation}
\delta E = \delta U + \delta P 
\end{equation} 
Pairing correction is in general smaller in amplitude and in antiphase with
shell correction; it has an effect of smoothing and reducing the total shell
plus pairing correction energy.  The experience of using Strutinsky's
method, gained by several nuclear scientists (e.g.  S.  Bj{\o}rnholm), was
also successfully employed to study shell effects in atomic cluster physics
and nanotechnology.

The inertia tensor \cite{bra72rmp} is given by
\begin{equation}
B_{ij} =2\hbar^2\sum_{\nu \mu} \frac{\langle \nu|\partial H/\partial
\beta_i|\mu \rangle \langle \mu|\partial H/\partial \beta_j|\nu
\rangle}{(E_\nu +E_\mu)^3}(u_\nu v_\mu +u_\mu v_\nu)^2  
\label{eq3}
\end{equation}
where $H$ is the single-particle Hamiltonian allowing to determine the
energy levels and the wave functions $|\nu \rangle$; $u_\nu^2$, $v_\nu^2$
are the BCS occupation probabilities, $E_\nu$ is the quasiparticle energy,
and $\beta_i, \beta_j$ are the independent shape coordinates.

For spherical fragments with $R, R_2$ deformation parameters the cranking
inertia symmetrical tensor will have three components, hence the scalar
\begin{equation}
B(R) =  B_{R_2R_2}\left( \frac{dR_2}{dR} \right)^2+
2B_{R_2 R}\frac{dR_2}{dR} + B_{RR}  
\label{br}
\end{equation}
or $ B= B_{22} + B_{21} + B_{11}$.
When we find the least action trajectory in the plane $(R,R_2)$ we need to
calculate the three components $B_{22}, B_{21}, B_{11}$ in every point of a
grid of 66$\times$24 (for graphics) or 412$\times$24 (for the real
calculation) for 66 or 412 values of $(R-R_i)/(R_t-R_i)$ and 24 values of
$\eta = (A_1-A_2)/A$ or $R_{2f}$.

\begin{figure}
\centerline{\includegraphics[width=8cm]{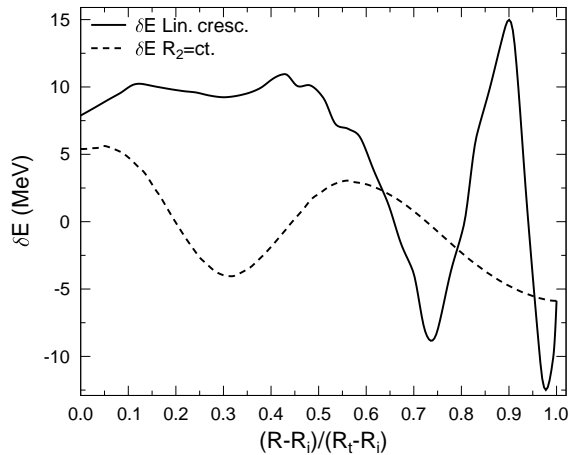}} %str252cf02lcbw}}
\caption{Comparison of shell plus pairing effects for fission 
of $^{252}$Cf with linearly increasing $R_2$ and constant $R_2$.
\label{cf02lc}} 
\end{figure}

We compare in figure \ref{shpe} the absolute values of shell and pairing
correction energies for symmetrical fission of $^{282}$Cn with
$R_2$~constant (dashed line) and linearly increasing $R_2$ (solid line).  As
expected, the gap for protons, $\Delta_p$, and neutrons, $\Delta_n$,
solutions of the BCS system of two equations, in figure~\ref{bcs} are also
following similar variations. Deep minima around $(R-R_i)/(R_t-R_i)=0.82$
are clearly seen in both figures. Similar results are also obtained for
heavy nuclei like $^{252}$Cf (see figure~\ref{cf02lc}) or $^{240}$Pu.
At the touching point, $R=R_t$, both kinds of variations of $R_2=R_2(R)$ are
ariving at the same state, hence the shell effects are identical there, as
may be seen in figures~\ref{shpe} and \ref{cf02lc}. 

For minimization of action we need not only $B_{RR}$ but also the values of
$B_{R_2R_2}, B_{R_2R}$ in every point of the above mentioned grid.  As
expected we obtained a dynamical path very different from the statical one. 
We could reproduce the experimental value of $^{282}$Cn spontaneous fission
half-life, $\log_{10} T_f^{exp} (s) =-3.086$.

In conclusion, with our method of calculating the spontaneous fission
half-life including macroscopic-microscopic method for deformation energy
based on asymmetric two-center shell model, and the cranking inertia for the
dynamical part, we may find a sequence of several trajectories one of which
gives the least action.  Assuming spherical shapes, we found that the shape
parametrization with linearly increasing $R_2$ is more suitable to describe
the fission process of SHs in comparison with that of exponentially or
linearly decreasing law.  It is in agreement with the microscopic finding
for $\alpha$~decay and cluster radioactivity concerning the preformation of
a cluster at the surface, which then penetrates by quantum tunneling the
potential barrier.

\begin{acknowledgments} 

This work was supported within the IDEI Programme under Contracts No. 
43/05.10.2011 and 42/05.10.2011 with UEFISCDI, and NUCLEU Programme
PN16420101/2016 Bucharest.  

\end{acknowledgments}

%\bibliographystyle{prsty}
%\bibliography{bibmycro,news,alpha,bibalfab,superheavy}

\end{document}